\documentclass[11pt,twoside]{article}

\usepackage{asp2006}
\usepackage{epsf}
\usepackage{psfig}
\usepackage{lscape}
\usepackage{graphicx}

\begin{document}
\title{AKARI and Spitzer observations of heavily 
obscured C-rich AGB/post-AGB stars}   
\author{D. A. Garc\'{\i}a-Hern\'andez$^1$, F. Bunzel$^2$, D. Engels$^2$,   
	J. V. Perea-Calder\'on$^3$, P. Garc\'{\i}a-Lario$^3$}   
\affil{
$^1$ Instituto de Astrof\'{\i}sica de Canarias. La Laguna, Tenerife, Spain \\
$^2$ Hamburger Sternwarte. Hamburg, Germany \\
$^3$ European Space Astronomy Centre, ESAC, ESA. Madrid, Spain
}    

\begin{abstract} 
We present AKARI/IRC and Spitzer/IRS observations of a selected sample of galactic
IRAS sources considered to be heavily obscured AGB/post-AGB stars based on their
characteristic IRAS colours. All of them are completely invisible in the optical
range but extremely bright in the infrared. Based on AKARI and Spitzer spectroscopy
and using DUSTY we are able to determine the dominant chemistry of their
circumstellar shells as well as the properties of the dust grains contained in these
shells. Most of the sources are found to be C-rich (being the reddest C-rich stars
observed so far). We find only molecular absorptions (and no PAH features) such as
acetylene (C$_2$H$_2$) at 13.7 $\mu$m, indicative of an early post-AGB stage. We
shortly discuss our findings in the context of stellar evolution during the hidden
``transition phase" from AGB stars to Planetary Nebulae.

\end{abstract}


\section{Motivation} 
Stars with low- and intermediate main-sequence masses (1$-$8 M$_{\odot}$) evolve
along the Asymptotic Giant Branch (AGB) just before they become Planetary
Nebulae (PNe). At the end of the AGB phase, these stars experience thermal
pulses (TP) and strong mass loss (up to 10$^{-4}$$-$10$^{-5}$
M$_{\odot}$yr$^{-1}$). As a consequence of the strong mass loss, the more
massive AGB stars (M$>$3 M$_{\odot}$) experience a hidden ``transition phase"
(AGB $=>$PN), in which the surrounding circumstellar envelope (CSE) of dust and
gas is optically thick, making these stars completely undetectable in the
optical domain but very bright in the IR. Thus, the departure from the AGB
occurs while these stars are still heavily obscured by their thick CSEs. In
addition, a fraction of this heavily obscured AGB population will be converted
into C-rich stars (M$<$4$-$5 M$_{\odot}$ at z$=$0), when the 3$^{rd}$ dredge-up
is able to bring freshly produced material to the stellar surface. More massive
AGB stars (M$>$4$-$5 M$_{\odot}$) will evolve as O-rich (the so-called OH/IR
stars) because of the ``Hot Bottom Burning" activation, which prevents the
carbon star formation (e.g. Garc\'{\i}a-Hern\'andez et al. 2007).

From the observational point of view, TP-AGB stars are found as Mira-like
variables with periods of up to 2000 days in the case of OH/IR stars
(Jim\'enez-Esteban et al. 2006) and up to 900 days in the case of ``extreme"
carbon stars (Kerschbaum et al. 2006). The strong mass loss and pulsation cease
when the AGB phase is terminated, and early post-AGB stars are characterized by
their non-variability status. Candidate post-AGB stars can be easily identified
by their characteristic IRAS colours and the IRAS two-colour diagram (see Fig.1)
is a powerful tool to discriminate between these candidate post-AGB stars and
other types of astronomical sources. Thus, heavily obscured post-AGB stars
display IRAS (25$-$12) $\geq$ 0.0 (those regions marked as IIIb, IV and V in
Fig.1). The obscured ``non-variable OH/IR stars" have been positively identified
as O-rich post-AGB stars (e.g. Engels 2002). This study was prompted by the
remarkable fact that a positive identification for the obscured C-rich post-AGB
candidates is still lacking. 

\begin{figure}[!ht]
\begin{center}
   \resizebox{0.6\hsize}{!}{
     \includegraphics*[0,50][810,572]{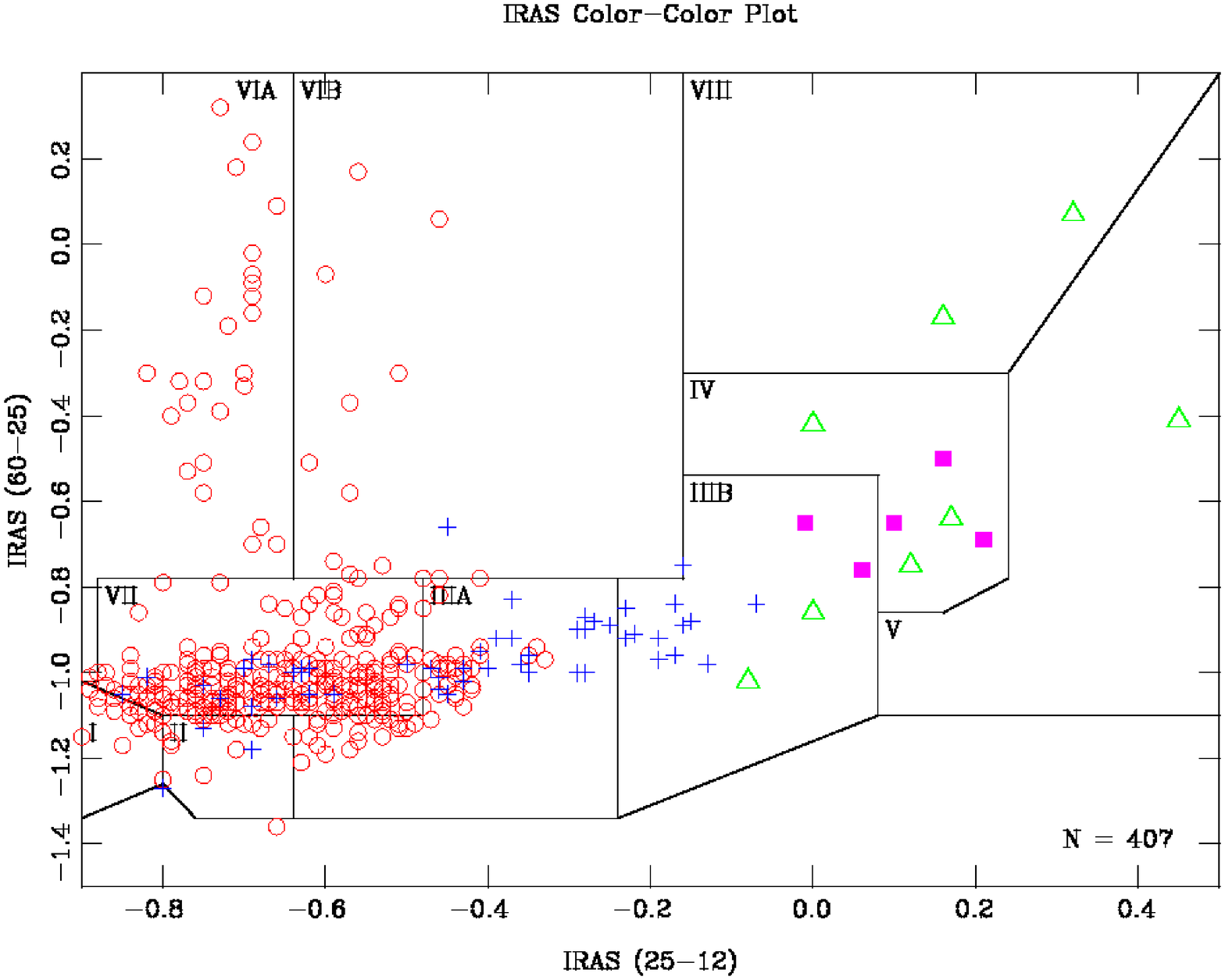}
    }
\end{center}    
\caption{IRAS two-colour diagram displaying the ``infrared carbon stars"
compiled from the literature by Chen \& Shan (2007) (red open circles), the
compilation of ``extreme carbon stars" from Chen \& Shan (2008) (blue crosses),
our heavily obscured C-rich AGB/post-AGB stars (magenta squares), and the sample
of obscured OH/IR sources (green open triangles) observed also
by us with AKARI (Bunzel et al., these proceedings).}
\label{fig:1}
\end{figure}

\section{The sample and infrared observations} 
Our sample is composed of 5 extremely red IRAS sources identified as candidate
post-AGB stars with no optical counterpart taken from the so-called GLMP
catalogue (see e.g. Garc\'{\i}a-Lario et al. 1997). These sources were of
unknown chemistry and with no detection of OH maser emission, and thus they were
expected to be heavily obscured C-rich stars. We observed our sample stars by
using the IRC and IRS spectrographs onboard AKARI and Spitzer satellites,
respectively.

The AKARI/IRC observations were carried out from 2006-12-08 until 2007-08-14. We
obtained long and short spectroscopic exposures by using the observation mode
AOT04 for different dispersion elements (NP, SG1, SG2), which give a spectral
resolution between 50 and 140 and an almost complete coverage from 1.8 to 12.9
$\mu$m (in one case we used LG2 that covers the 17.5$-$25.7 $\mu$m region). The
data reduction was done with the IRC Spectroscopy Toolkit Version 20080528. A
few sources were reduced by applying the short exposure mode of the data
reduction pipeline, since all long exposure frames were saturated.

The Spitzer spectral data were taken with IRS under a General Observer
program ($\#$30258, PI, P. Garc\'{\i}a-Lario). Spitzer/IRS spectra of
our sample stars were obtained by using the Short-High (SH: 9.9$-$19.6
$\mu$m; R$\sim$600) module. The Spitzer spectra were reduced with the
help of the Spitzer IRS Custom Extractor (SPICE) and SMART (see
e.g. Garc\'{\i}a-Hern\'andez et al. 2009, in press).

\begin{figure}[!ht]
\begin{center}
      \includegraphics*[width=6.5cm,angle=-90]{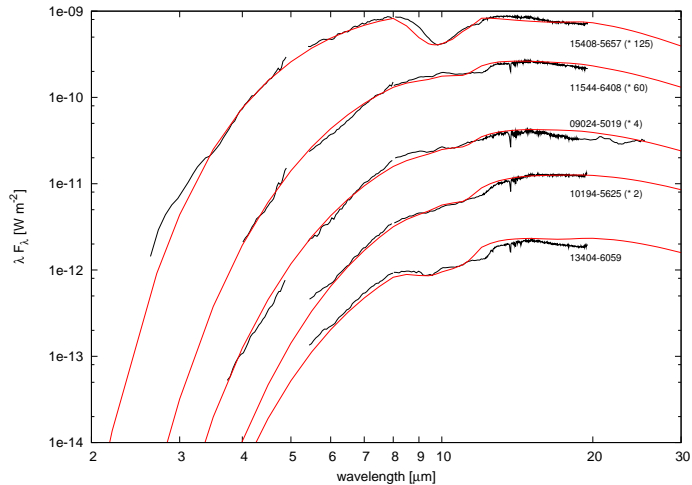}
\end{center}    
\caption{AKARI/IRC and Spitzer/IRS spectra (in black) with the fitted DUSTY
models overimposed (in red) of our sample sources. Note the molecular
absorptions at 13.7 $\mu$m due to acetylene (C$_2$H$_2$).
}
\label{fig:2}
\end{figure}

\section{SEDs and DUSTY modelling} 
Spectral energy distributions (SEDs) of all 5 sources were constructed with the
reduced AKARI/IRC and Spitzer/IRS spectra (see Fig.2). For all sources (except
IRAS15408) we obtained almost featureless spectra (e.g. typical of carbonaceous
grains) with a weak 9$-$12 $\mu$m depression (which might be attributed to small
amounts of SiC or amorphous silicates) and acetylene (C$_2$H$_2$  at 13.7
$\mu$m) absorptions (Fig.2). IRAS15408 is a particular case and it shows the 10
$\mu$m absorption feature of amorphous silicates (O-rich) and no C$_2$H$_2$
absorption.

We have determined the dominant chemistry and the properties of the dust grains
present in the CSE by fitting DUSTY models (Ivezi\'c \& Elitzur 1995) to the
observed SEDs. The best fits obtained are shown in Fig.2 and the derived model
parameters from DUSTY are summarized in Table 1\footnote{We fixed the
temperature of the central star (T$_{eff}$=2500 K), the condensation temperature
of the dust (T$_{dust}$=1000 K), and the grain size (a=0.27 $\mu$m) because the
spectral shapes of the models are not very sensitive to variations of these
parameters.}. Our modelling confirms the C-rich nature of 4 out of 5 sample
stars, containing $>$ 80\% of carbonaceous dust. The only exception is IRAS15408
for which we found almost half amorphous silicate and half amorphous carbon
dust.

\begin{table*}
\small
\caption{Derived model parameters (optical depth at 10 $\mu$m and abundances of
amorphous carbon, amorphous silicates and SiC) from DUSTY.}          
\label{dusty}      
\begin{center}
\begin{tabular}{r c c c c}		
\hline 
\hline 
IRAS name    & $\tau$(10$\mu$m)  &       & Abundances       &   \\ 
             &                   & n$_C$ & n$_{Sil}$        &  n$_{SiC}$ \\
             &                   & [\%]  & [\%]             & [\%]  \\
\hline
09024$-$5019 & 7   & 90 & 5 & 5  \\
10194$-$5625 & 9   & 90 & 5 & 5  \\
11544$-$6408 & 6   & 86 & 7 & 7  \\
13404$-$6059 & 10   & 80 & 15 & 5  \\
15408$-$5657 & 8   & 48 & 52 & 0  \\
\hline                  
\end{tabular}
\end{center}
\end{table*}

\section{Discussion} 

It is not completely clear, if these heavily obscured C-rich stars are still
pulsating on the AGB or if they have already departed from the AGB, being the
C-rich counterparts of the heavily obscured ``non-variable OH/IR stars" (O-rich 
post-AGB stars). The only variability information comes from the IRAS variability index and with the exception of IRAS09024 (with a high 99\% probability for variability), all sources are probably not large amplitude
variables (with $<$20\% probability for variability), and thus we identifiy them
as heavily obscured C-rich post-AGB stars. 

From Fig.1, it is obvious that these sources are the reddest AGB/post-AGB stars
studied so far (being even redder than the ``extreme" carbon stars previously
known). Interestingly, IRAS09024 (probably the unique AGB star in our sample)
shows the strongest C$_2$H$_2$ absorption (Fig.2). C$_2$H$_2$ and other carbon-based
molecules are believed to be the
building blocks of more complex molecules such as PAHs (e.g. Sloan et al. 2009),
which are seen in more evolved (and less obscured) C-rich post-AGB stars and PNe
of our Galaxy. Thus, the detection of C$_2$H$_2$ in our ``non-variable extreme carbon
stars" is indicative of an early post-AGB stage because PAHs have not formed
yet. We are probably witnessing the evolution of the more massive and obscured
C-rich AGB/post-AGB stars of our Galaxy.


\acknowledgements 
This research is based on observations with AKARI and Spitzer, a JAXA project
with the participation of ESA and a NASA's Great Observatories Program,
respectively.


\end{document}